# Nuclear deformation effect on the level statistics

A. Al-Sayed
*Physics Department, Faculty of Science, Zagazig University, Zagazig, Egypt*

**Abstract**

We analyze the nearest neighbor spacing distributions of low-lying $2^+$ levels of even-even nuclei. We grouped the nuclei into classes defined by the quadrupole deformation parameter ($\beta_2$). We calculate the nearest neighbor spacing distributions for each class. Then, we determine the chaoticity parameter for each class with the help of the Bayesian inference method. We compare these distributions to a formula that describes the transition to chaos by varying a tuning parameter. This parameter appears to depend in a non-trivial way on the nuclear deformation, and takes small values indicating regularity in strongly deformed nuclei and especially in those having an oblate deformation.

## 1. Introduction

There are two main ways to investigate nuclear structure; one of them is the detailed study of individual nuclei. This method tests different nuclear models, and gives us the internal structure of the nucleus under investigation. However, this method does not allow constructing a correlation relation between different nuclei. In addition, we are in need to investigate a large number of studies to understand nearly 2500 discovered nuclei. The second way is the cumulative study, in which one attempts to classify nuclei in terms of a small number of parameters that can be related to nuclear structure. This method finds the correlations between different nuclei and helps us to understand the nuclear structure evolution through the nuclear chart.

Several papers have been published [see, e.g. 1,2,3,4,5], whose purpose is to find the extent of chaos in nuclear dynamics. Random matrix theory (RMT) has gained a great success in the description of the fluctuation properties of spectra of quantum systems. It models a chaotic system by an ensemble of random matrices subject only to symmetry constraints. Systems conserving time reversal, such as the atomic nucleus are described by the Gaussian orthogonal ensemble (GOE) [6].

The nearest neighbor spacing (NNS) distribution occupies an important place in the statistical theory of spectra. Unfortunately, RMT does not provide a closed form expression for the NNS distribution, which is suitable for the analysis of experimental data. However, Wigner proposed an approximate expression, which is exact in the case of 2 x 2 matrices. The NNS distribution of nuclear spectra at neutron threshold energies is in good agreement with the Wigner distribution [7]. The study of level statistics at low-lying excitation energies requires complete (few or no missing levels) and pure (few or no unknown spin-parities) level schemes. Unluckily complete and pure level schemes are only available for a limited number of nuclides. Therefore, to improve the statistical significance of the study, spacing distributions for several sequences of levels (each sequence has the same spin parity) from different nuclides have been combined.

Such statistical studies concluded that the NNS distribution of nuclei in the ground state region is intermediate between the Poisson distribution expected for regular systems and the Wigner distribution for chaotic ones. In addition, light nuclides showed a behavior close to chaotic near the ground state, while heavier nuclides seemed to be more regular. This can be understood from the fact that the fluctuation properties of the spectra given by the nuclear shell model (which well describes light nuclei) agree with the GOE.

In the present study, we investigate the effect of nuclear deformation -as a measure of nuclear structure- on the chaoticity of even-even nuclei. Due to serious statistical limitation of the experimental data basis, earlier results yielded only qualitative information concerning the effect of deformation. Some efforts have focused on the symmetries of the IBA, but again the statistics were too limited to reach definite conclusions. In this paper we wish to establish a direct relation between deformation and nuclear level statistics by parameterizing the available experimental data via quadrupole deformation parameter.

We focus on the $2^+$ states because of their abundance in even-even nuclei. The data set and the deformation parameter used to classify the nuclei are described in Sec. 2. Section 3 describes the technique of the analysis, while the results are given in Sec. 4.

## 2. Data Set

In this section, we describe our choice of levels in even-even nuclei. We recall that the collective motion is interpreted as vibrations and rotations of the nuclear surface in the geometric collective model first proposed by Bohr and Mottelson [8], where the nucleus is modeled as a charged liquid drop. The moving nuclear surface may be described by an expansion in spherical harmonics with time-dependent shape parameters as coefficient. The quadrupole deformation seems to be the most important collective excitations of the nucleus. For axially symmetric nuclei, the nuclear radius can be written as,

$$R(\theta,\phi) = R_{av}[1+\beta_2 Y_{20}(\theta,\phi)], \tag{2}$$

where $\beta_2$ is the quadrupole deformation parameter. Positive and negative $\beta_2$ values correspond to prolate and oblate shapes respectively. The deformation parameter suffers from the difficulty to distinguish between static deformation and the dynamic amplitude of the quadrupole vibration in soft spherical nuclei. It can be obtained only by a detailed analysis of spectra and transition probabilities. Despite this difficulty, we draw a crude picture of the dependence of nuclear level statistics on nuclear deformation in a direct way.

The quadrupole deformation parameters $\beta_2$ are taken from macroscopic-microscopic calculations [9]. In order to obtain a sufficient number of energy levels within each interval, we take the absolute value of $\beta_2$. The $Z$ dependence of deformation is not taken into account. We compare results by analyzing the data set in terms of the experimental $\beta_2$ values, as deduced from the $B(E2;\, 0^+ \rightarrow 2^+)$ values [10].

The data on low-lying $2^+$ levels of even-even nuclei are taken from Endt [11] for $22 \leq A \leq 44$, and from the Nuclear Data Sheets until March 2005 for heavier nuclei. We consider nuclei in which the spin-parity $J^\pi$ assignments of at least five consecutive levels are definite. In cases where the spin-parity assignments are uncertain and where the most probable value appeared in brackets, we accept this value. We terminate the sequence in each nucleus when we arrive at a level with unassigned $J^\pi$, or when an ambiguous assignment involved a spin-parity among several possibilities, as e.g. $J^\pi = (2^+, 4^+)$. We chose one of the suggested assignments, when only one such level occurred in the sequence, and was followed by several definitely assigned levels containing at least two levels of the same spin-parity, provided that the ambiguous level is found in a similar position in the spectrum of a neighboring nucleus. However, this situation has occurred for less than 5% of the levels considered. In this way, we obtained 1132 levels of spin-parity $2^+$ belonging to 150 nuclei.

## 3. Method of Analysis

For statistical studies using RMT, one should have a spectrum of unit mean level spacing. This is obtained by fitting a theoretical expression to the number $N(E)$ of levels below excitation energy $E$; this process is called unfolding. The expression used here is the constant-temperature formula (4),

$$N(E) = N_0 + \exp\left(\frac{E - E_0}{T}\right). \tag{4}$$

The three parameters $N_0$, $E_0$ and $T$ obtained for each nucleus vary considerably with mass number. Nevertheless, all three show a clear tendency to decrease with increasing mass number. A detailed account of the method of analysis used in the present work has been given in Refs. [12, 13], and references therein.

The nuclear states are characterized by their invariance under time reversal and space rotation, which can be represented by the Gaussian orthogonal ensemble (GOE) of random matrices. The NNS distribution of levels of the GOE is well approximated by Wigner's distribution [7]

$$Pw(s) = \frac{\pi}{2} s \exp\left(-\frac{\pi}{4} s^2\right). \tag{5}$$

Here, $s$ is the spacing of neighboring levels in units of the mean level spacing. For integrable systems, the NNS distribution is generically given by the Poisson distribution,

$$P_P(s) = \exp(-s). \tag{6}$$

The key ingredient of the present analysis is the assumption that the deviation of the NNS distribution of low-lying nuclear levels from the GOE statistics is caused by the neglect of possibly existing conserved quantum numbers other than energy, spin, and parity. A given sequence $S$ of levels can then be represented as a superposition of $m$ independent sequences $S_j$ each having fractional level density $f_j$, with $j = 1, \ldots, m,$ and with $0 < f_j \le 1$ and $\sum_{j=1}^{m} f_j = 1$. We assume that the NNS distribution $P_j(s)$ of $S_j$ obeys GOE statistics. The exact NNS distribution $P(s)$ of this superposition has been given in Ref. [6]. It depends on the $(m-1)$ parameters $f_j$, $j = 1, \ldots, m-1$. In [14], this expression has been simplified by observing that $P(s)$ is mainly determined by short-range level correlations. This reduces the number of parameters to unity and the proposed NNS distribution of the spectrum is

$$P(s, f) = \left[1 - f + f(0.7 + 0.3f)\frac{\pi s}{2}\right] \times \exp\left(-(1 - f)s - f(0.7 + 0.3f)\frac{\pi s^2}{4}\right), \tag{7}$$

which depends on only one parameter, the mean fractional level number $f = \sum_{j=1}^{m} f_j^2$ for the superimposed sub-spectra. This quantity will eventually be used as a fit parameter. For a large number $m$ of sub-spectra, $f$ is of the order of $1/m$. In this limit, $P(s, f)$ approaches the Poisson distribution $P(s, 0) = P_P(s)$. This expresses the well-known fact that the superposition of many GOE level-sequences produces a Poissonian sequence. On the other hand, when $f \to 1$ the spectrum approaches the GOE behavior. Indeed, $P(s,\ 1)$ coincides with the Wigner distribution (5), this is expected as the system then consists of a single GOE sequence. This is why $f$ is called the chaoticity parameter.

In this study, we use the parameter $f$ instead of empirical Brody's interpolation formula $P(s,\omega)=\omega(\gamma+1)s^{\gamma}\exp(-\omega s^{\gamma+1})$, where $\gamma$ is a fitting parameter and $\omega=\left[\Gamma\left(\frac{\gamma+2}{\gamma+1}\right)\right]^{\gamma+1}$ [15]. Since, formula (7) has a theoretical basis because $f$ represents the area of phase space of the wave function occupied by the chaotic dynamics. We plot a relation in figure 1 between the variation of parameter $f$ and the parameter $\omega$, by calculating the chi-square fit between the NNS distribution in Eq. (7), and Brody distribution. $\chi^2=\int_0^5\left(P(s,f)-P(s,\omega)\right)^2 ds$. By defining the parameter $f$, we calculate the $\omega$ parameter. We observe nearly a linear relationship between the two parameters, that strength our choice of parameter $f$.

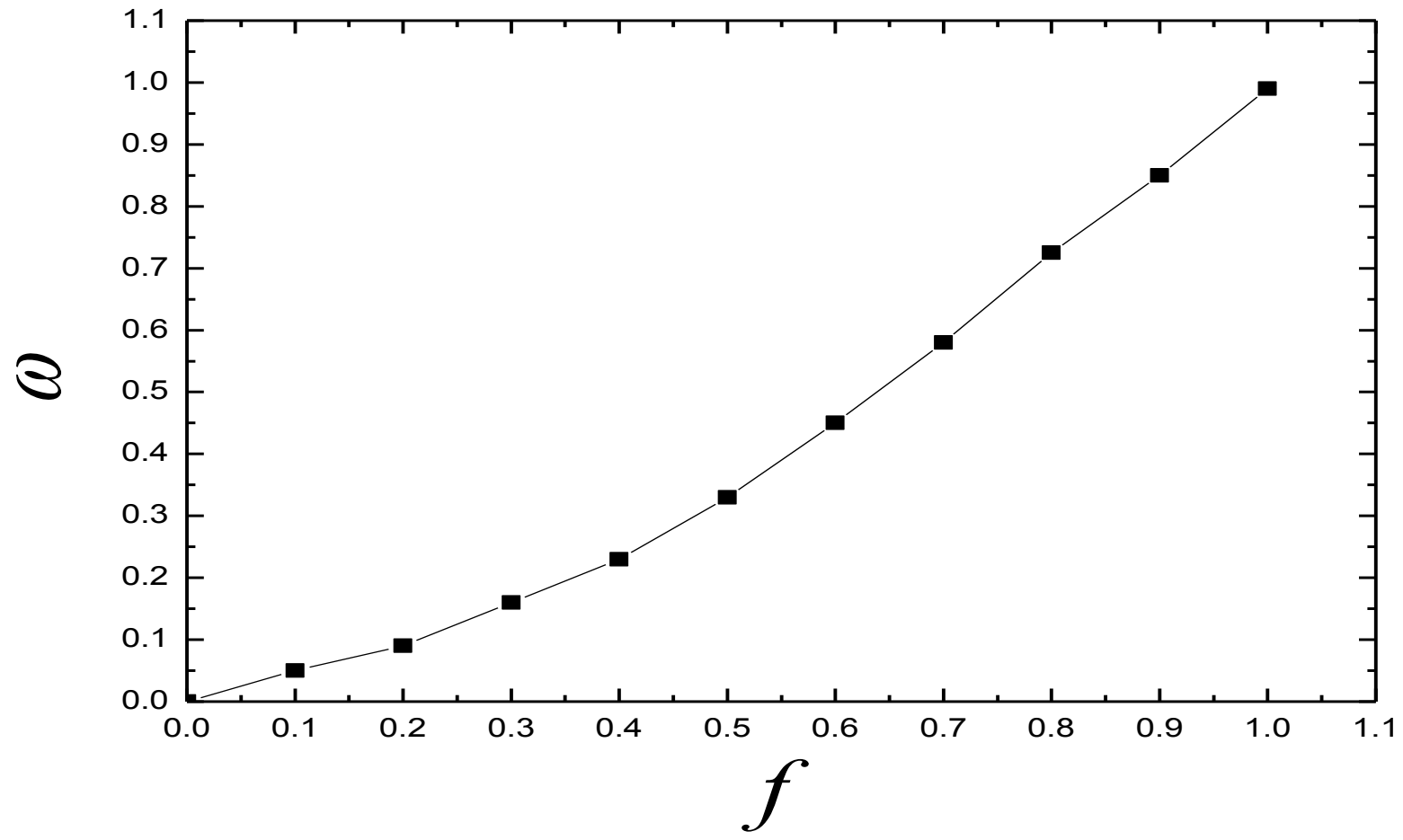

Fig. 1: A relation between a parameter $f$ and $\omega$ of Brody distribution.

A common method to determine the parameter $f$ is via the best fit obtained by chi-square. But working with histograms of experimental data, there is always the issue of selection of bin size. Different choices of the bin width, particularly when the widths vary, can lead to quite different visual conclusions. In order to avoid this difficulty, Bayesian inference [16] is used to determine the parameter $f$. The Bayesian analysis is independent of the bin size; it deals with the spacings directly [13]. The Bayesian analysis yields the best-fit value of the chaoticity parameter $f$ and its error for each NNS distribution. For a given sequence of spacings $s = (s_1, s_2, \ldots s_N)$, the joint probability distribution $P(s|f)$ of these spacings, conditioned by the parameter $f$, is given by

$$P(s\,|f)=\prod_{i=1}^{N}P(s_i,f),\qquad (8)$$

with $P(s_i, f)$ given by Eq. (7). Bayes' theorem provides the posterior distribution

$$P(f\,|s)=\frac{P(s\,|f)\mu(f)}{M(s)},\qquad (9)$$

of the parameter $f$ given the events $s$. Here, $\mu(f)$ is the prior distribution and

$$M(s) = \int_0^1 P(s \mid f)\mu(f)\,\mathrm{d}f\,, \tag{10}$$

is the normalization.
The prior distribution is found from Jeffreys' rule [17]

$$\mu(f) \propto \left| \int P(s \mid f)[\partial \ln P(s \mid f)/\partial f]^2 \,\mathrm{d}s \right|^{1/2}. \tag{11}$$

By evaluating numerically the prior distribution (11), and substituting Eq. (8) into Eq. (11), and approximating the result by the polynomial of sixth order in *f*. We obtain

$$\mu(f) = 1.975 - 10.07f + 48.96f^2 - 135.6f^3 + 205.6f^4 - 158.6f^5 + 48.63f^6. \tag{12}$$

The distribution $P(s|f)$ takes very small values even for only moderately large values of $N$. Because of this fact, the accurate calculation of the posterior distribution becomes a formidable task. In order to simplify the calculation, Eq. (8) may be rewritten in the form

$$P(s \mid f) = \exp\left(-N\phi(f)\right), \tag{13}$$

where

$$\phi(f) = (1-f)\langle s\rangle + \frac{\pi}{4}f(0.7+0.3f)\langle s^2\rangle - \left\langle \ln\left[1-f+\frac{\pi}{2}f(0.7+0.3f)s\right]\right\rangle. \tag{14}$$

Here, the notation

$$\langle x\rangle = \frac{1}{N}\sum_{i=1}^{N} x_i\,, \tag{15}$$

has been used. One finds the function $\phi(f)$ to have a deep minimum, say at $f = f_0$. One can therefore represent the numerical results in analytical form by parameterizing $\phi$ as

$$\phi(f) = A + B(f-f_0)^2 + C(f-f_0)^3, \tag{16}$$

where the parameters $A, B, C$ and $f_0$ are implicitly defined by (14).
We then obtain

$$P(f \mid s) = c\mu(f)\exp\left(-N\left[B(f-f_0)^2 + C(f-f_0)^3\right]\right), \tag{17}$$

where $c = \exp(-NA)/M(s)$ is the new normalization constant. When $P(f|s)$ is not Gaussian, the best-fit value of $f$ cannot be taken as the most probable value. Rather we take the best-fit value to be the mean value $\bar{f}$ and measure the error by the standard deviation $\sigma$ of the posterior distribution, i.e.

$$\bar{f} = \int_0^1 fP(f \mid s)\,\mathrm{d}f\,, \text{ and } \sigma^2 = \int_0^1 (f-\bar{f})^2 P(f \mid s)\,\mathrm{d}f\,. \tag{18}$$

The chaoticity parameter $\bar{f}$ and the standard deviation $\sigma$ are calculated for each group of nuclei. The results are given in the following section.

## 4. Results

The search for a phenomenological "control parameter" to describe the evolution of the stochastic nature of nuclear dynamics became a subject of nuclear structure in the last two decades.

In the present work, we examine the use of the quadrupole deformation parameter as probe of nuclear structure. In spite of absence of complete theoretical study to define a certain fixed values of $\beta_2$ corresponding to critical points of shape/structural transitional regions, we tend to classify nuclei into fixed intervals of $\beta_2$ to get a qualitative study. We recall that the analysis of many short sequences of levels tends to overestimate the degree of chaoticity measured by a parameter *f.* We focus our attention not on the absolute values of *f* but on the way *f* changes with $\beta_2$.

I follow the same method of analysis given in Ref [12]. By grouping nuclei according to their $\beta_2$ deformation parameter, instead of the ratio $R_{4/2}$ (the ratio between lowest $4^+$ to $2^+$ level states). The work done in this paper may be a thankless work if there is a simple linear relationship between $\beta_2$ and $R_{4/2}$. But on plotting a direct plot between them we get figure 2, which shows almost uncorrelated relation. We start to classify the available even-even nuclei by $\beta_2$ to find what can we get? Although the capability of the usage of $R_{4/2}$ to identify the collective excitation of nuclei has been confirmed in many published papers depending on both theoretical e.g. [18, 19] and empirical [20, 21] calculations. But the $R_{4/2}$ didn't introduce an acceptable method to differentiate between deformed nuclei whether they are oblate or prolate deformed, and hence their degree of chaoticity, while the $\beta_2$ parameter do so.

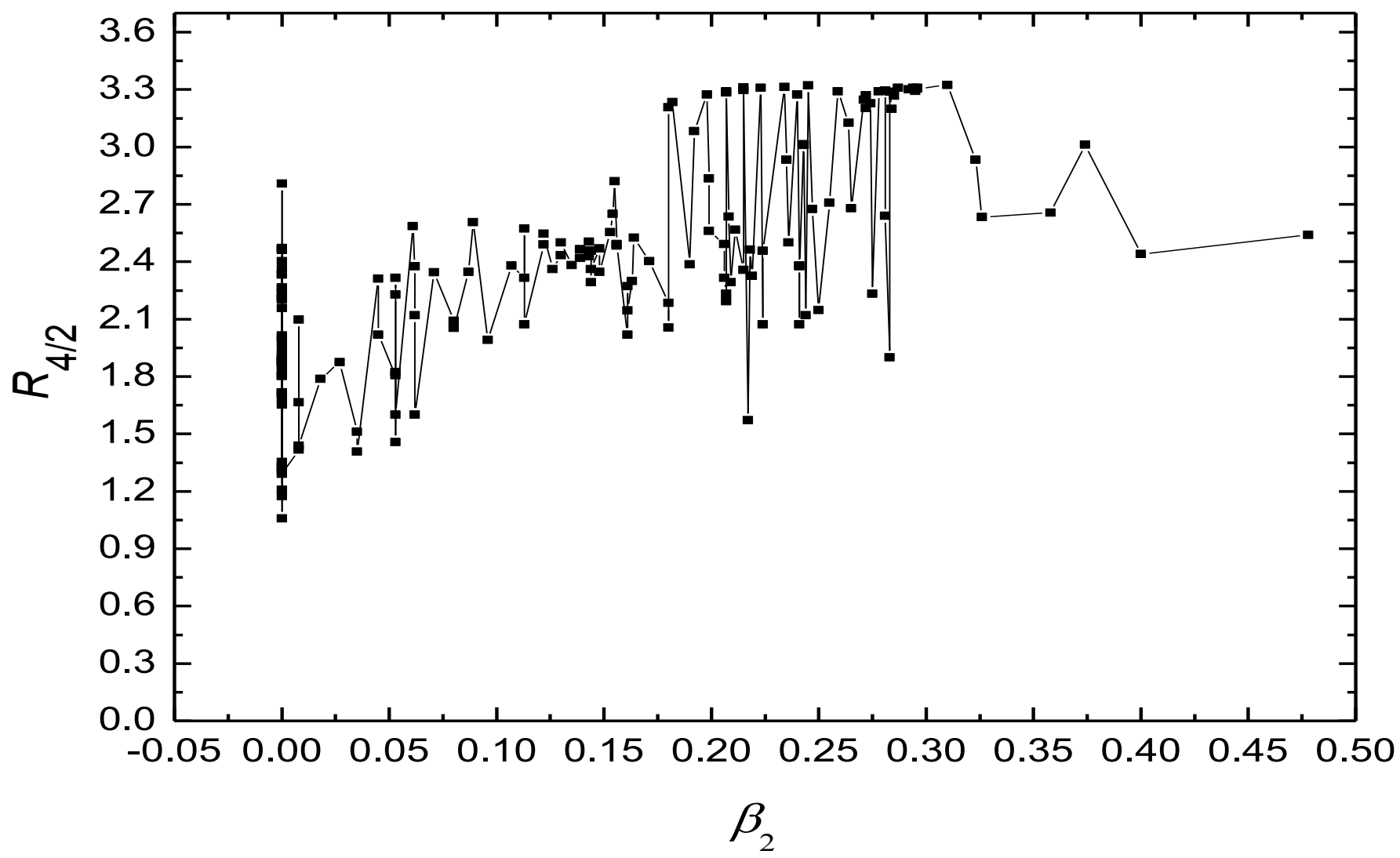

Fig. 2: A plot between $R_{4/2}$ and quadrupole deformation parameter $\beta_2$.

Figure 3 shows a comparison of the spacing distributions that depends on the parameter *f* and the histograms for nuclei divided into classes according to the deformation parameter $\beta_2$. In view of the small number of spacings within each class, the agreement seems satisfactory. By plotting the best-fit values obtained by Bayesian analysis of the chaoticity parameters against $\beta_2$, we get figure 4. From this figure we see the apparent chaoticity of nuclei having $\beta_2$ equal or nearly equal to zero these are spherical (magic or semi magic) nuclei which are expected to have shell model spectra, thus agree with GOE.

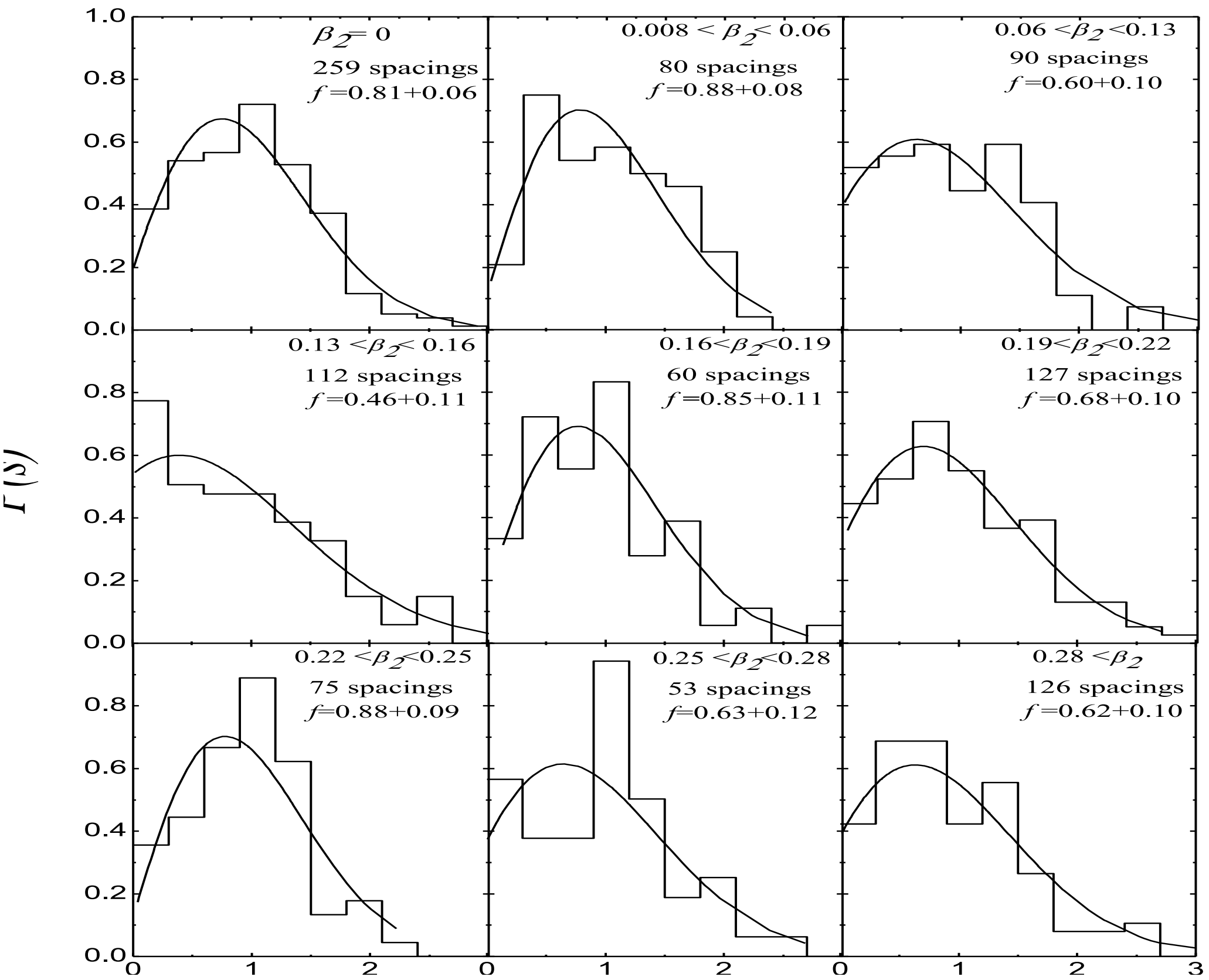

Fig. 3: NNS distribution of even-even nuclei classified according to deformation parameter $\beta_2$

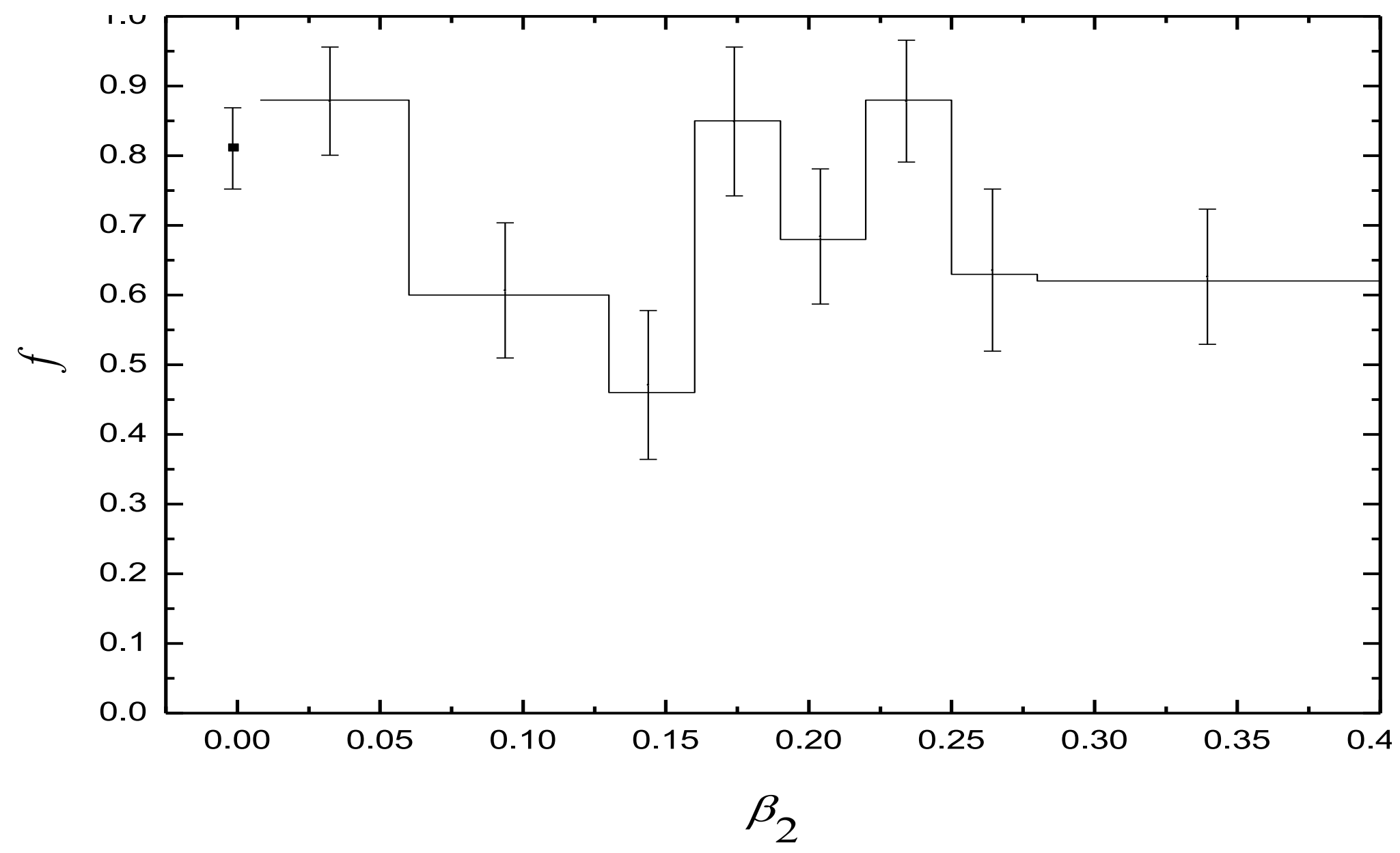

Fig. 4: The chaoticity parameter $f$ and standard deviation $\sigma$ (error bars) against theoretical deformation parameter $\beta_2$.

An observed minimum of chaoticity is centered at $\beta_2 \approx 0.20$. Although the statistical errors are not good enough to draw a conclusion, but it may indicate that something interesting happening through that interval. Two remarkable minima are nearly centered at $\beta_2 \approx 0.26$ and $\beta_2 \approx 0.34$, these two intervals have nearly the same chaoticity parameter that may refer to well deformed nuclei.

Finally, there is a minimum at $\beta_2 \approx 0.14$, in order to understand this regular behavior we will not study this interval individually, since we observe also a regular behavior through its neighbor interval centered at $\beta_2 \approx 0.095$. As these two intervals are statistically significant from their neighbors, we will extend our study to cover the two intervals from $\beta_2 \approx 0.06$ to $\beta_2 \approx 0.16$. By a detailed study of nuclei contributed to that interval, we notice nearly an equal number of prolate-oblate nuclei. So, which kind is the responsible for this regular behavior? We neglected the negative sign of deformation of oblate nuclei, but now the role to test the correctness of such proposal comes to surface. We obtained 110 levels of spin-parity $2^+$ belonging to 15 nuclei of oblate shape $^{62}$Ni, $^{124}$Te, $^{140}$Sm, $^{190}$Pt, $^{192}$Pt, $^{194}$Pt, $^{196}$Pt, $^{198}$Pt, $^{200}$Pt, $^{192}$Hg, $^{196}$Hg, $^{198}$Hg, $^{200}$Hg, $^{202}$Hg, and $^{204}$Hg while 123 levels belonging to 16 nuclei of prolate shape $^{82}$Kr, $^{84}$Kr, $^{88}$Kr, $^{76}$Ge, $^{80}$Se, $^{82}$Se, $^{92}$Sr, $^{96}$Mo, $^{106}$Cd, $^{108}$Cd, $^{110}$Cd, $^{112}$Cd, $^{132}$Ba, $^{136}$Ce, $^{144}$Ce, and $^{192}$Os. The NNS distribution is given in figure 5; it is interesting to find that the oblate deformed nuclei have more regular spectra than prolate one. This may help us to understand the apparent regularity. Indeed, we show [22] that this is not only a special case to that interval, but it is the general rule, that oblate nuclei are more regular than prolate ones. This fact may be interpreted as, the degree of interaction between single particle motion which is chaotic and collective motion of whole nucleons which believed to be more regular is weaker in case of oblate deformed nuclei than prolate ones.

Finally, we compare our results based on theoretically calculated $\beta_2$, with those of experimental $\beta_2$ values, as deduced from the $B(E2;\ 0^+ \rightarrow 2^+)$ values [10]. Two disadvantages of using experimental $\beta_2$ arise. One of them is that all nuclei having non-zero values, so, it is difficult to differentiate between oblate-prolate deformed nuclei. On the other hand, $\beta_2$ values are not available for all nuclei under investigation in this study. We obtained 1050 levels of spin-parity $2^+$ belonging to 137 nuclei. Figure 6 shows the trend of dependence of chaoticity parameter on experimental $\beta_2$. In spite, the error bars are statistical insignificant, the figure shows nearly the same trend observed in figure 4.

## 5. Conclusion

We study the nearest neighbor spacing distribution of even-even nuclei classified according to the quadrupole deformation parameter using the available experimental energy levels. We use a model that interpolates between the Poisson (regular) to Wigner (chaotic) distribution by varying the chaoticity parameter from 0 to 1 respectively. The observed regular behavior of nuclei having certain values of deformation parameter indicate that there must be a lot of work to qualify the usage of such parameter as an indicator to variation in nuclear structure, and this work may be visualized as a step on the road.

## Acknowledgement

I am grateful to Prof. A. Y. Abul-Magd, Faculty of Science, Zagazig University, for his helpful discussions and scientific support.

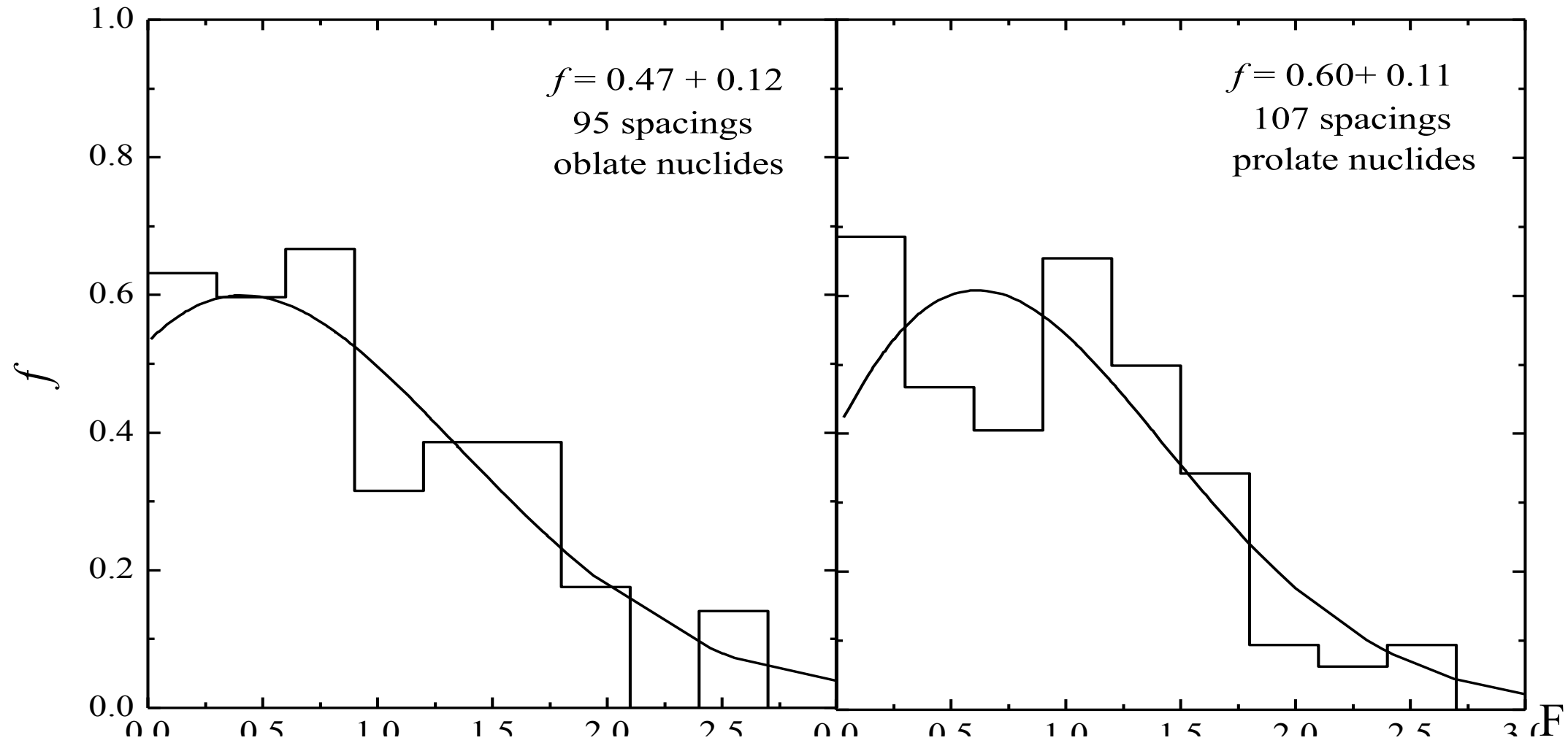

Fig. 5: Chaoticity of oblate-prolate deformed even-even nuclei in the range of $\beta_2$ = 0.06 to 0.16

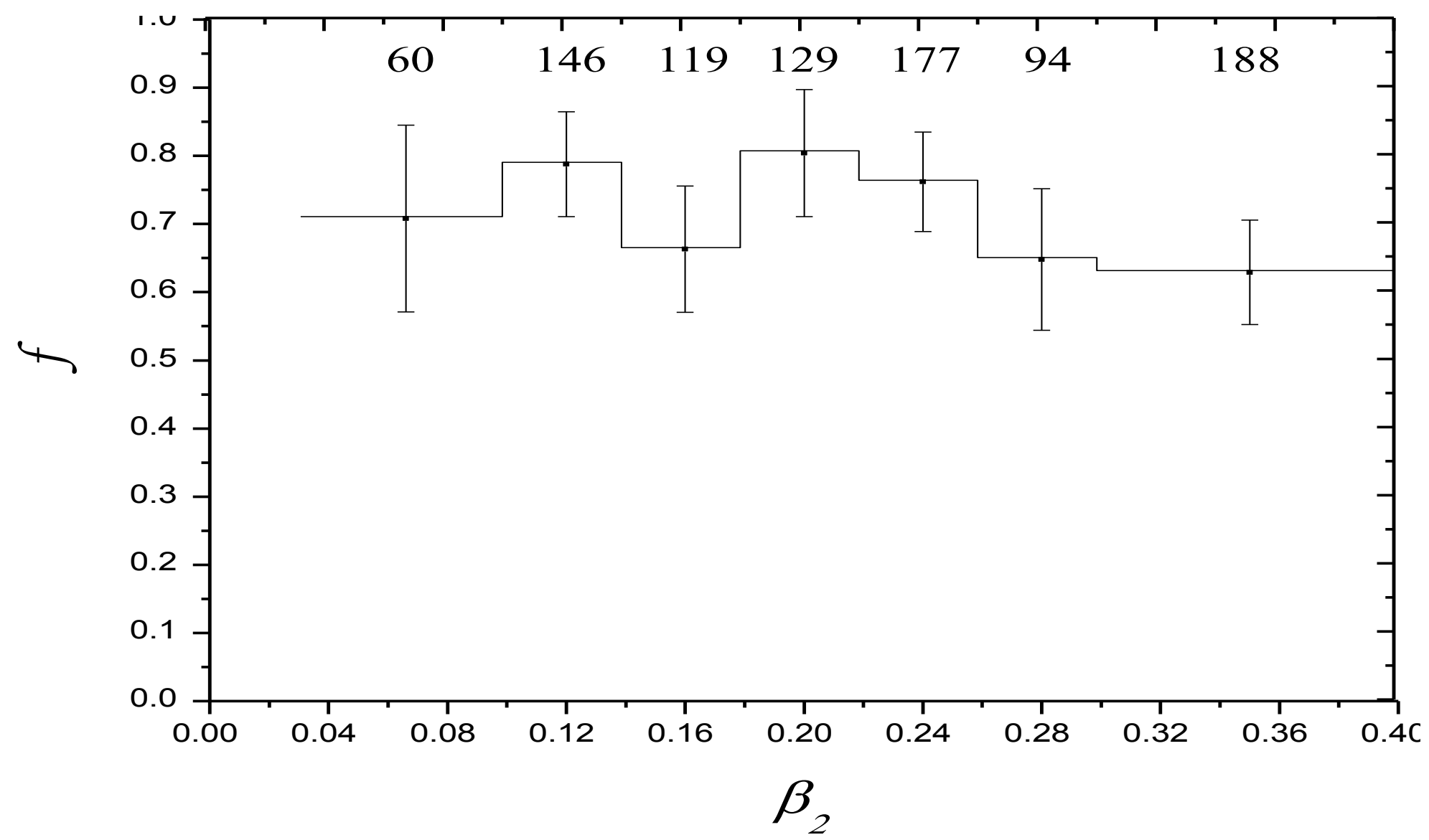

Fig. 6: The chaoticity parameter $f$ and standard deviation $\sigma$ (error bars) against experimental deformation parameter $\beta_2$, the number of spacings is given above each ratio.